\documentclass{article}

\usepackage{epsfig}
\usepackage{amsmath}
\usepackage{mathtools}
\usepackage{amssymb}
\usepackage{color}
\usepackage{graphicx}
\graphicspath{ {./Figures/} }
\usepackage{todonotes}
\usepackage{comment}
\usepackage{arxiv}
\usepackage[utf8]{inputenc} 
\usepackage{fontenc}    
\usepackage{hyperref}       
\usepackage{url}            
\usepackage{booktabs}       
\usepackage{amsfonts}       
\usepackage{nicefrac}       
\usepackage{microtype}      

\usepackage{graphicx}
\usepackage{amssymb}
\usepackage{verbatim}
\usepackage{tikz}
\usetikzlibrary{arrows,automata}
\usepackage{algpseudocode}
\usepackage{algorithm}
\usepackage{amsmath}
\usepackage{mathtools}
\usepackage{todonotes}

\usetikzlibrary{arrows.meta}
\tikzset{>={Latex[width=3mm,length=3mm]}}

\usepackage{amsthm}
\theoremstyle{definition}

\theoremstyle{plain}

\theoremstyle{remark}

\title{Improving Run Length Encoding by Preprocessing}
\author{Sven Fiergolla \textnormal{and} Petra Wolf\\
	Abteilung Informatikwissenschaften -
	Fachbereich 4\\
	Universität Trier -
	Germany\\
	\texttt{\{fiergolla, wolfp\}@informatik.uni-trier.de}}
\begin{document}
	\maketitle
	%

%
%
%
%
%
%
%
%
%

\begin{abstract}
	The Run Length Encoding (RLE) compression method is a long standing simple lossless compression scheme which is easy to implement and achieves a good compression on input data which contains repeating consecutive symbols. In its pure form RLE is not applicable on natural text or other input data with short sequences of identical symbols. We present a combination of preprocessing steps that turn arbitrary input data in a byte-wise encoding into a bit-string which is highly suitable for RLE compression. The main idea is to first read all most significant bits of the input byte-string, followed by the second most significant bit, and so on. We combine this approach by a dynamic byte remapping as well as a Burrows-Wheeler-Scott transform on a byte level. Finally, we apply a Huffman Encoding on the output of the bit-wise RLE encoding to allow for more dynamic lengths of code words encoding runs of the RLE. With our technique we can achieve a lossless average compression which is better than the standard RLE compression by a factor of 8 on average.
%
\keywords{Lossless data compression \and Vertical byte reading \and Run length encoding \and Burrows-Wheeler-Scott transform \and Dynamic byte remapping \and Huffman encoding}
\end{abstract}

\section{Introduction}
In the last decades, digital data transfer became available everywhere and to everyone. This rise of digital data urges the need for data compression techniques or improvements on existing ones. \emph{Run Length Encoding}~\cite{rle-patent} (abbreviated as RLE) is a simple coding scheme that performs lossless data compression. 
It identifies each maximal sequence of consecutive identical symbols of a string by a \emph{run}, usually denoted by $\sigma^i$, where $\sigma$ is an alphabet symbol and $i$ is its number of repetitions. 
%
To give an example, the string \emph{aaaabbaaabbbb} consists of the four runs
$a^{4}b^{2}a^{3}b^{4}$. 
In the standard RLE compression scheme the number of bits $\ell$ reserved to encode the length of a run is fixed. Each run is encoded by $\ell$ bits storing the binary representation of the length of the run, followed by the binary encoding of the letter of the run (which usually also has some fixed length $d$).
Some strings like \emph{aaaabbbb} achieve a very good compression rate because the string contains only two different characters which repeat more than twice. 
Hence, for $\ell = 8$ and $d=8$, its \emph{RLE-representation} $a^4b^4$ can be stored in 4 bytes, instead of 8 bytes needed for the original string in \emph{ASCII} or \emph{UTF-8}.
On the other hand, if the input consists of highly mixed characters with few or no repetitions at all like \emph{abababab}, the RLE-representation of the string is $a^1b^1a^1b^1a^1b^1a^1b^1$ which needs 16 bytes for $\ell=d=8$. 
Thanks to its simplicity RLE is still being used in several areas like fax transmission, where RLE compression is combined with other techniques into \emph{Modified Huffman Coding} \cite{fax-rle} and applied on binary images. As most fax documents are typically simple texts on a white background~\cite{palette-image}, RLE compression is particularly suitable for fax and often achieves good compression ratios.

But RLE also has a major downside, which is obviously the possible explosion in size, due to missing repetitions in the input string. Expanding the string to twice the original size is a rather undesirable worst case behavior for a compression algorithm, so one has to make sure the input data is fitted for RLE as compression scheme.
In this work, we present a combination of preprocessing techniques that increases the average compression ratio of the RLE compression scheme on arbitrary input data. The main idea is to consider a bit-wise representation of the data and to read all bits in a row, which have the same position in a byte. We combine this approach with dynamic byte remapping and a Burrows-Wheeler-Scott transform (BWST for short) to increase the average run length on a bit level.
We show experimentally that with the help of such preprocessing the originally proposed RLE can compress arbitrary files of different corpora. 
Our proposed algorithm is even comparable to the popular compression scheme \texttt{ZIP}.
Files suited for regular RLE are compressed even more than with the original method.
To unify the measurements, the relative file size after compression is calculated by encoding all files listed in the Canterbury and Silesia Corpus individually. 
Since most improvements like permutations on the input, for example, a reversible BWST to increase the number of consecutive symbols or a different way of reading the byte stream, take quite some time, encoding and decoding speed will decrease with increasing preprocessing effort compared to regular RLE.

This work is structured as follows.
In the next section, we discuss the literature on RLE after giving some preliminaries. Then, we discuss our proposed technique in more detail and evaluate it in comparison with the standard RLE compression scheme and \texttt{ZIP} v3.0 afterwards.



\section{Preliminaries}
Throughout this work, we assume $\Sigma$ to be a finite alphabet. A string $w$ is a sequence $c_1,...,c_n$ of letters $c_i \in \Sigma$, $1 \leq i \leq n$. The set of all such sequences is denoted by $\Sigma^*$ which is the free monoid over $\Sigma$, with concatenation as operation and with the empty word $\varepsilon$ as neutral element. In standard text representation, the string $w$ is coded as an array $S_w$ of $n$ blocks of bit-strings, each of size $8$, that can be read and written at arbitrary positions, and where the $i$-th block of $S_w$ contains the binary representation of the $i$-th character of $w$. In the following, our algorithm is working on a \emph{byte alphabet}, i.e., 8 bits are assumed to encode one input symbol. For the examples discussed later this byte alphabet is realized as an UTF-8 encoding.
 The \emph{vertical interpretation}, also called \emph{Bit-Layers text representation} in~\cite{DBLP:conf/sofsem/CantoneFS20}, codes the array $S_w$ as an ordered collection of $8$ binary strings of length $n$, $(B_7, B_6, . . . , B_0)$, where the $j$-th binary string $B_j$ is the sequence of bits
at position $j$ of the blocks in $S_w$ encoding characters in $w$, in the order in which they appear in $w$, where $j=0$ refers to the least significant bit.
Let $\chi \colon \Sigma^* \to \{0,1\}^*$ define a compression scheme. For a string $w \in \Sigma^*$ let $m_w$ be the number of bytes in the UTF-8 encoding of $w$. We define the number of \emph{bits per symbol (bps)} of $w$ under $\chi$ as $\frac{|\chi(w)|}{m_w}$.

\section{Combination with other compression methods}
	Examples of combining different techniques to achieve a better compression rate has already been discussed in other papers and achieved good compression ratios, not much worse than the theoretical limit of around 1.5 \textit{bps} \cite{kolmogorov},
	for example, Burrows and Wheeler used their transform, in combination with a Move-to-Front Coder and a Huffman Coder~\cite{Burrows94}.
	Also standard compression algorithms, such as \texttt{bzip2}~\cite{bzip2-link} use a combinations of transforms, i.e., by default \texttt{bzip2} applies a RLE, a Burrows-Wheeler Transform followed by a Huffman encoding. Via parameters it is also possible to enable a second run length encoding on the character level between the latter two phases. In contrast to our approach, both RLEs are on a sequential horizontal byte level and not on a vertical binary level. 
	
	Generally, a combined approach would no longer be considered preprocessing but it clearly has some benefits over the encoding of regular RLE runs with a fixed size. 
	The fax transmission implementation also uses RLE and Huffman coding together~\cite{fax-rle}. While the idea of encoding the RLE runs with Huffman codes is already known and analyzed~\cite{rle-patent}, it is mostly in a static sense and optimized for special purpose applications such as fax transmission and DNA sequences \cite{rle-bio,rle-dna}. However, the vertical byte reading enables new approaches, even more in combination with the idea of byte remapping and becomes applicable to more than just binary fax or DNA sequences, with longer runs of any kind in average. As our evaluation shows, our technique makes nearly every type of input data suitable to RLE.

\section{Proposed technique}

\begin{figure}
	\includegraphics[width=\textwidth]{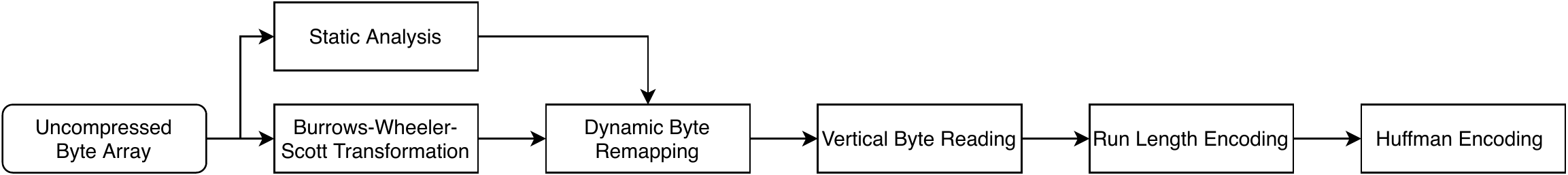}
	\caption{Schematic illustration of the proposed algorithm.}
	\label{fig:Scheme}
\end{figure}
The binary representation of an arbitrary string does not contain long runs of repeating bits, but, first, reading all most significant bits of all bytes, then all second most significant bits and so on, results in much longer average runs of the same bit value. This is partially explained by ASCII artifacts but also by the small Hamming distance of the binary string representations of most letters, as they all have a value between 65 and 122 in the UTF-8 encoding. This improvement in average run length
can even be enhanced by mapping the byte values of the input to lower values in relation to their occurrence probability. To further improve the algorithm we use a variable length code for encoding the RLE runs instead of a fixed size. This way, the proposed algorithm can compress arbitrary files with a reasonable compression ratio and even improve against regular RLE on files highly suited for the original algorithm.

The proposed technique is depicted in Figure~\ref{fig:Scheme}. 
In the first step, the uncompressed byte array is analyzed and for each byte its number of occurrences is counted. In parallel, a bijective  Burrows-Wheeler-Scott Transform~\cite{DBLP:journals/corr/abs-1201-3077} is applied to the input byte array, which produces a reversible permutation of the input byte array with long repetitions of similar symbols. Afterwards, each byte is remapped, where the most frequent byte values are mapped to the lowest binary values. The resulting byte array is then interpreted in a specific way, at first all most significant bits of all bytes are read, then all second most significant bits and so on, resulting in long average runs of identical bits. On this representation, a run length encoding is applied and the runs are counted to generate a Huffman tree. Using this, the runs are output with a variable length code, together with the relevant mapping needed to decompress the file.
Next, we discuss each step of our proposed technique in detail. We will sequentially apply each step to the example input string $S=abraca$.
The binary UTF-8 interpretation of the example string $S = abraca$ contains
3 runs of length 3 and 4, 9 runs of length 2 as well as 8 runs of length 1 in total.
$$
bin(S) = 01100001\ 01100010\ 01110010\ 01100001\ 01100011\ 01100001 
$$

\subsection{Burrows-Wheeler-Scott Transform}
Working with arbitrary data implies starting with an uncompressed byte array, which is analyzed by the static analysis component. All occurrences of each byte value are counted and later on used in the byte mapping process. In the mean time, a Burrows-Wheeler-Scott transform \cite{DBLP:journals/corr/abs-1201-3077} (BWST for short) is performed on the same uncompressed byte array, using the C library LibDivSufSort \cite{LibDivSufSort}. The BWST is a strong enhancement of the classical Burrows Wheeler Transformation (BWT)(introduced in~\cite{Burrows94} and analyzed in~\cite{manzini2001analysis}) which is used in a variety of compression algorithms. 
In short, the BWT creates all cyclic permutations of the input strings and sorts them lexicographically.  
As the last symbol of a cyclic permutation is the predecessor of the first symbol, in the last column of all permutation, identical symbols are clustered together if the input string contains repetitions, e.g., repeating natural words.
Then, the last column of all permutations in this sorting is output.
So, in general, the BWT increases the number of subsequent identical symbols.

Here, we use the enhanced BWST transform, which
in contrast to the original BWT
does not require additional information, nor start and stop symbols or the index of the original permutation in the sorting. Briefly, it does not construct a matrix of all cyclic rotations like the originally proposed BWT, instead it is computed with a suffix array sorted with DivSufSort, closer described in~\cite{DBLP:conf/stringology/0001K17} and~\cite{inducedSortingSA}. This allows transforming the input in its entirety, rather than splitting it into small chunks.
Since we do not alter the BWST algorithm and only use an existing library as a building block in our preprocessing pipeline, we refer for more algorithmic details on BWST to~\cite{DBLP:journals/corr/abs-1201-3077}. In Theory it should be possible to perform a BWT in linear time and space~\cite{Burrows-linear-time}, but at the time being, there is no implementation available and the java library kanzi~\cite{kanzi} is used.
Applying BWST on the input string $S=abraca$ results in the string $caraab$ with the binary representation
$$S_{\text{BWST}} = 01100011\ 01100010\ 01110010\ 01100001\ 01100001\ 01100001.$$

\subsection{Dynamic Byte Remapping}
	Next, we apply a dynamic byte remapping of the input data, where the most frequently used bytes are mapped to the lowest value.
	This way the values are not alternating in the whole range of 0 to 255 and between 65 and 122 for regular text, but rather in a smaller subset and the most frequent ones will be the smallest values.
	Hence, considering only the most significant bits of each byte, the number of consecutive zeros should increase, yielding longer average runs of RLE on a vertical byte reading.
	Let $\Sigma$ be the set of all bytes appearing in the input data. Then, let $p^* \colon \Sigma \to \{0, 1\}^8$ be the function applying the dynamic byte remapping. 
%
	Considering our example string $S_{\text{BWST}}=caraab$, the most frequent letter is $a$, followed by an $b, c, r$ which appear once each. 
	By fixing an order on $\{b, c, r\}$ we get the byte remapping function $p^*(a) = 00000000$, $p^*(b) = 00000001$, $p^*(c)= 00000010$, and $p^*(r)=00000011$.
	Applying $p^*$ on $S_{\text{BWST}}$ yields $$S_{\text{DBR}}=00000010\ 00000000\ 00000011\ 00000000\ 00000000\ 00000001$$
	
	
	For huge input files, splitting the input and creating a single map for each block of data should result in lower average values used but also creates some kind of overhead because the mapping has to be stored in the encoded file as well. Applying a single mapping to lower values for the whole file still results in increased runs in the vertically interpreted bytes and is used in our approach.

\subsection{Vertical Byte Reading}
Reading all most significant bits of all bytes, then the second most significant bits of all bytes and so on greatly improves the average run length on a bit level for most types of files as shown in the example below.

Recall the binary UTF-8 interpretation of the example string $S = abraca$ as $bin(S) = 01100001\ 01100010\ 01110010\ 01100001\ 01100011\ 01100001$
with 3 runs of length 3 and 4, 9 runs of length 2 as well as 8 runs of length 1 in total.
The vertical byte reading codes the string $S$ as an ordered collection of 8 binary strings of length $l(S) = n$, where the $i$'th binary string $B_{i}$ is the sequence of bits at position $i$ of the bytes in $S$, in the order in which they appear in $S$, where $i=0$ refers to the least significant bit.
 We refer to the concatenated bit vectors $B_{7} \dots B_{0}$ induced by such representation as the \emph{vertical representation} of the encoding. Formally, letting $p(c)$ be the binary encoding, for $c \in \Sigma$ and assume therefore that $p(a) = 01100001$, $p(b) = 01100010$, $p(c) = 01100011$ and $p(r) = 01110010$. 
Hence, the vertical representation of the string $S$ is:
\begin{align*}
	B_7 &= 000000 & B_6 &= 111111 &
	B_5 &= 111111 & B_4 &= 001000\\
	B_3 &= 000000 & B_2 &= 000000 &
	B_1 &= 011010 & B_0 &= 100111 	
\end{align*}
Performing RLE on the consecutive bits of $B_7$ to $B_0$ results in 5 runs of length 6, 2 runs of length 3, 3 runs of length 2 and just 6 runs of length 1 as opposed to the many short runs of the simple interpretation. This is because the binary similarity between the used characters, as the character for $a$ and $b$ only differ in one bit. It is clear that simply a different way of reading the input does not compress the actual data, instead it enables a better application of existing compression methods. This approach can also be generalized to arbitrary sized alphabets. By shrinking the alphabet to actually used code words, the numbers of bit vectors can be reduced which is discussed in~\cite{DBLP:conf/sofsem/CantoneFS20}.

Now, let us continue with our toy example and apply the vertical byte reading to the string $S_{\text{DBR}}$ from the last step. This gives us the vertical representation $S_{\text{VBR}} = B_7B_6 \dots B_0$ which highlights in contrast with the above vertical representation of the initial string $S$ the impact of the dynamic byte remapping step.
\begin{align*}
	B_7 &= 000000 & B_6 &= 000000 &
	B_5 &= 000000 & B_4 &= 000000	\\
	B_3 &= 000000 & B_2 &= 000000 &
	B_1 &= 101000 & B_0 &= 001001	
\end{align*}

\subsection{Run Length Encoding}
 Continuing with the example, and performing RLE on the consecutive bits of $B_7\dots B_0$ results in 1 run of length 36, 1 of length 5, 1 of length 2, and 5 runs of length 1.
In general the binary RLE simply counts alternating runs of ones and zeros and encodes the length of the run into a fixed length code with $n$ bits \cite{rle-patent}. Assuming a run always starts with a zero and the maximum run length $2^n -1$ determined by the length of the code, we add an artificial run of length 0 to flag a run exceeding the maximum run length or a run starting with 1. This way we can encode any binary string. Some experiments with different default maximum run lengths showed improvement in performance but also revealed some shortcomings. Refining the algorithm to use different maximum run lengths for the different bit vectors $B_{7}$,$ B_{6}$, $\dots$ , $B_{0}$ did improve but not solve the issue of being a very static solution. It is possible to choose maximum run lengths to work more efficient for a specific file or to be adequate for a range of files but it is always a trade off. Eventually, a variable length code for encoding the runs was needed, so the algorithm is combined with another compression method, namely Huffman Encoding. The maximum run length is limited to 255, in order to limit the size of the Huffman tree and therefore the average prefix length. This gives us the RLE representation $S_{\text{RLE}} = \langle \gamma_0,\gamma_1, \dots, \gamma_k \rangle$ with $ k \in \mathbb{N}$ and $ \gamma \in [0,255]$.
$$S_{\text{RLE}} = \langle 36, 1, 1, 1, 5, 1, 2, 1 \rangle$$

\subsection{Huffman Encoding of RLE runs}
While the RLE is performed with a fixed maximum run length set to 255 to limit the amount of Huffman codes to be generated, the occurrence of each run is counted. After the RLE step is finished, a Huffman tree for the runs is generated \cite{huffman} and each run is encoded with the according prefix free code of variable length. This further reduces the required space to encode the file but also a representation of the Huffman tree needs to be persisted to reverse the variable length coding. For ease of decoding, a map of run length to the pair of length of prefix, and prefix is generated. Finally, the size of the map, followed by the map is written to the stream. The Huffman tree for the runs of $S_{\text{RLE}}$ generates the following prefixes $1 \rightarrow 0,  2 \rightarrow 10,5 \rightarrow 110, 36 \rightarrow 111$, which encodes $S_{\text{RLE}}$ to the final encoded output $S_{\text{Huf}}$ with 13 bits:
$$S_{\text{HUF}} =  111 \ 0 \ 0 \ 0 \ 110 \ 0 \ 10 \ 0.$$

\section{Decoding}
The decoding happens in three phases. First, the size of the byte map is parsed to know how many pairs of bytes are expected. 
In the second phase, the map of Huffman prefixes is parsed and the number of expected pairs is determined.
Since each pair in the Huffman map consists of the byte which is mapped, the length of the prefix and the prefix itself, it is easy to decode each mapping from the stream. After both required maps are parsed, the compressed content follows. The following stream is read bit-wise to be able to match any bit sequence of variable length to the related Huffman code and decode it into a number of runs. Reversing RLE from the decoded runs recreates the bit vectors $B_7 \dots B_0$ which are written to the output file. Finally, the byte mapping parsed in phase 1 is applied to the file and the bijective BWST is inverted, restoring the original input data.
\section{Evaluation}

\begin{table}[b]
	\centering
	\begin{tabular}{l|r|r|l|l}	
		file & file size [kB] & $\Sigma$ size & type & description\\
		\hline
		dickens &9956 & 100 & English text & Collected works of Charles Dicken \\
		mozilla & 50020 & 256 & exe & Tarred executables of Mozilla\\
		mr & 9740 & 256 & picture & Medical magnetic resonance image  \\
		nci & 32768 & 62 & database & Chemical database of structures \\
		ooffice & 6008 & 256 & exe & A dll from Open Office.org 1.01  \\
		osdb & 9852 & 256 & database & Sample database in MySQL \\
		reymont & 6472 & 256 & pdf & Polish text  \\
		samba & 21100 & 256 & exe & Tarred source code of Samba 2-2.3\\
		sao & 7084 & 256 & bin & The SAO star catalog  \\
		webster & 40488 & 98 & htaml & The 1913 Webster Unabridged Dictionary\\
		xml & 5220 & 104 & xml & Collected XML files  \\
		xray & 8276 & 256 & picture & X-ray medical picture  \\
	\end{tabular}
	\caption{The Silesia Corpus.}
	\label{tab:t05 The Silesia Corpus}
\end{table}	

\begin{table}[b!]
	\centering
	\begin{tabular}{l|r|r|r|r|r|r|r|r}	
		file & \multicolumn{1}{c|}{original}  & \multicolumn{3}{c|}{RLE} & \multicolumn{3}{c|}{proposed algorithm} &  impr. \\ 
		& \multicolumn{1}{c|}{size [kB]} &  s. [kB]& r.s.\ [\%] & [\textit{bps}] & s. [kB]& r.s.\ [\%] & [\textit{bps}] & [\%] \\
		\hline
		alice29.txt & 152.1 & 604.9 & 397.70 & 31.82 & 65.4 & 43.00 & 3.44 & 89.19\\
		asyoulik.txt & 125.2 & 514.8 & 411.18 & 32.90 & 59.2 & 47.28 & 3.79 & 88.50\\
		cp.html & 24.6 & 98.9 & 402.03 & 32.16 & 11.0 & 44.72 & 3.60 & 88.88\\
		fields.c & 11.2 & 44.6 & 398.21 & 32.01 & 5.1 & 45.54 & 3.72 & 88.57\\
		grammar.lsp & 3.7 & 14.8 & 400.00 & 31.89 & 1.9 & 51.35 & 4.13 & 87.16\\
		kennedy.xls & 1029.8 & 1820.3 & 176.76 & 14.14 & 229.8 & 22.32 & 1.79 & 87.38\\
		lcet10.txt & 426.8 & 1749.7 & 409,96 & 32.80 & 170.5 & 39.95 & 3.20 & 90.26\\
		plrabn12.txt & 481.9 & 1944.9 & 403.59 & 32.29 & 215.6 & 44.74 & 3.58 & 88.92\\
		ptt5 & 513.2 & 136.6 & 26.62 & 2.12 & 82.1 & 16.00 & 1.28 & 39.90\\
		sum & 38.2 &  99.4 & 260.21 & 20.80 & 19.6 & 51.31 & 4.10 & 80.28\\
		xargs.1 & 4.2 & 17.7 & 421.43 & 33.50 & 2.5 & 59.52 & 4.76 & 85.88\\
		\hline
		all files & 2811.9 & 7046.6 & 250.60 & 20.05 & 862.7 & 30.68 & 2.45 & 87.76\\
		\multicolumn{2}{l|}{$\varnothing$ values per file} & - & 337.06 & 26,95 & - & 42.34 & 3.40 & 83.18
	\end{tabular}
	\caption{The Canterbury Corpus encoded with RLE and the proposed algorithm. For each method, absolute file size in kB after compression, relative file size (size after compression)/(original size) in~\% and bps are listed. The last column shows the improvement of the proposed algorithm over RLE as $1-\,$(size proposed algorithm)/(size RLE) in~\%.}
	\label{tab:results_canterbury}
\end{table}
To evaluate the effectiveness of the proposed  technique, a collection of files from the Canterbury Corpus~\cite{CalgaryCorpusCritic}, and the Silesia Corpus (also containing medical data)~\cite{silesia} were compressed. A more detailed list of the files contained in the Silesia Corpus can be found in Table \ref{tab:t05 The Silesia Corpus}, all file sizes are given in kB (kilo byte). 
The relative file sizes after compression are listed in Tables~\ref{tab:results_canterbury} and \ref{tab:results_silesia}. To have another unit of measure, the \emph{bps} (bits per symbol) is also shown in the table. Plain RLE on a bit level with a maximum run length of 255, encoded in 8 bits, showed good results on the file \emph{ptt5}, a ITU-T standard fax showing a black and white picture. This fits our expectations since RLE was designed for those types of files. On this file, simple RLE achieved a relative file size of 26\% compared to the original size which relates to $2.1$ bits per symbol. 
On all files contained in the Canterbury corpora, the plain bit level RLE increases the files by a factor of $3.3$ on average.

%

\begin{table}[t!]
	\centering
	\begin{tabular}{l|r|r|r|r|r|r|r|r}	
		file & \multicolumn{1}{c|}{original}  & \multicolumn{3}{c|}{\texttt{ZIP}} & \multicolumn{3}{c|}{proposed algorithm} &  impr.\\ 
		& \multicolumn{1}{c|}{size [kB]} &  size [kB] & r.s.\ [\%] & [\textit{bps}] & size [kB] & r.s.\ [\%] & [\textit{bps}] & [\%]\\
		
		\hline
		dickens & 9956 & 3780 & 37.96 & 3.03 & 3964 & 39.82 & 3.19 & -4.87\\
		mozilla & 50020 & 18604 & 37.19 & 2.98 & 24808 & 49.60 & 3.97 & -33.35\\
		mr & 9740 & 3608 & 37.04 & 2.96 & 2908 & 29.86 & 2.39 & \textbf{19.40}\\
		nci & 32768 & 3128 & 9.55 & 0.76 & 2900 & 8.85 & 0.71 & \textbf{7.29}\\
		ooffice & 6008 & 3028 & 50.40 & 4.03 & 3904 & 64.98 & 5.20 & -28.93\\
		osdb & 9852 & 3656 & 37.11 & 2.97 & 3260 & 33.09 & 2.65 & \textbf{10.83}\\
		reymont & 6472 & 1816 & 28.06 & 2.25 & 1960 & 30.28 & 2.42 & -7.93\\
		samba & 21100 & 5336 & 25.29 & 2.02 & 7012 & 33.23 & 2.66 & -31.41\\
		sao & 7084 & 5208 & 73.52 & 5.88 & 5604 & 79.10 & 6.33 & -7.60\\
		webster & 40488 & 11920 & 29.44 & 2.36 & 12012 & 29.67 & 2.37 & -0.77\\
		xml & 5220 & 676 & 12.95 & 1.04 & 868 & 16.63 & 1.33 & -28.40\\
		xray & 8276 & 5900 & 71.29 & 5.70 & 4840 & 58.48 & 4.68 & \textbf{17.97}\\
		\hline
		all files & 206984 & 66660 & 32.21 & 2.58 &74040 & 35.77 & 2.86 & -11.07\\
		\multicolumn{2}{l|}{$\varnothing$ values per file} & - & 37.48 & 3.00 & - & 39.47 & 3.16 & -7.31
	\end{tabular}
	\caption{The Silesia Corpus encoded with \texttt{ZIP} v3.0 and the proposed algorithm. For each method, absolute file size in kB after compression, relative file size (size after compression)/(original size) in~\% and bps are listed. The last column shows the improvement (bold if $>0$) of the proposed algorithm over \texttt{ZIP} as $1-\,$(size proposed algorithm)/(size \texttt{ZIP}) in~\%.}
	\label{tab:results_silesia}
\end{table}

	In contrast, our presented technique, consisting of a combination of preprocessing steps and a Huffman encoding of the RLE runs, 
	achieved, with a relative file size of 40.8\% on average, comparable results to the state of the art for both corpora. Already suited files, like the file \emph{ptt5} from the Canterbury Corpus, were compressed even further than with plain bit level RLE. On the Pizza \& Chill Corpus~\cite{} the proposed algorithm achieves a compression of 36.4%
	For comparison, \texttt{ZIP} v3.0 using a combination of the dictionary technique LZ77 and Huffman codes, is listed. All zip compressions were executed with \texttt{zip -evr \$file}. For instance, \texttt{ZIP} achieves an average relative file size of 37.5\% on the single files in the Silesia Corpus, where our algorithm achieves 39.5\%.

\begin{table}[b!]
	\centering
	\begin{tabular}{l|r|r|r|r|r}	
		file type & \multicolumn{2}{c|}{\texttt{ZIP}} & \multicolumn{2}{c}{proposed algorithm} \\
		& $\varnothing$ rel.\ size [\%] & $\varnothing$  [\emph{bps}] & $\varnothing$  rel.\ size [\%] & $\varnothing$  [\emph{bps}] & improvement [\%]\\
		\hline
		PGM & 76.77 & 6.14 & 70.54 & 5.64 &\textbf{8.11}\\
		PPM & 76.37 & 6.11 & 70.13 & 5.61 &\textbf{8.17}\\
		DNG & 87.09 & 6.96 & 85.12 & 6.80 &\textbf{2.26}\\
		OBJ & 26.92 & 2.15 & 36.40 & 2.91 &-35.21\\
		STL & 43.93 & 3.51 & 64.30 & 5.14 &-46.36\\
		PLY & 33.88 & 2.71 & 41.87 & 3.34 &-23.58\\
	\end{tabular}
	\caption{Average relative file size after compression of a random selection of files of different file types compressed with \texttt{ZIP} v3.0 in comparison with the proposed algorithm. The last column shows the improvement (bold if $>0$) of the proposed algorithm over \texttt{ZIP} as $1-\,$(size proposed algorithm)/(size \texttt{ZIP}) in~\%.}
	\label{tab:results_zip}
\end{table}

In a second evaluation, a randomly chosen collection (listed in detail under~\cite{raw}) of raw image files and 3D-object files (taken from~\cite{shilane2004princeton}) were compressed with the proposed algorithm and with \texttt{ZIP} in version 3.0. The average relative file sizes are listed in Table~\ref{tab:results_zip}, all files were compressed individually. Regarding large raw picture files like .PPM and .PMG from the Rawzor corpus~\cite{rawzor} as well as a random collection of .DNG files from raw.pixel.us~\cite{raw}, a higher compression ratio than obtained by \texttt{ZIP} could be achieved. 3D-object files in the encoding format .obj .sty and .ply are also compressed by our algorithm to a size comparable but inferior to the output produced by \texttt{ZIP}. 
This shows that with our approach run length encoding can become a suitable compression algorithm for more than just pellet based images like fax transmissions.

\section{Implementation}
The implementation is hosted on Bitbucket and released under the MIT license. The source code and the test data can be found here \cite{own}. All source code is written in Kotlin and runs on any Java virtual machine, but performs best executed on the GraalVM \cite{vsipek2019exploring}.
%
%

All benchmark tests were performed on a system running Linux Pop OS with a 5.6.0 kernel with an AMD Ryzen 5 2600X six core processor (12 threads) with a 3.6 GHz base clock and a 4.2 GHz boost clock speed. For memory, 16GB 3200MHz ram and a Samsung evo ssd was used for persistent storage. 

	Encoding is reasonably fast with measured 7.1 seconds but the decoding is rather slow with 16.7 seconds for the Canterbury Corpus with a memory peak of  19396 kB.
	The much larger Silesia Corpus took about 16 minutes do compress and about an hour to decompress, which shows that the current implementation is not very optimized and cannot handle large files that well.
	Avoiding internal operations and large or complex data structures to hold all the input data or even collecting the values of same significance in memory into byte arrays greatly improved time performance of the algorithm described. It has to be mentioned that there is still some potential in performance optimization and parallelization. 
	In theory, all 8 runs could be created at the same time by reading the input as a byte stream which would vastly improve overall encoding speed instead of the currently used library to handle the binary stream~\cite{IoStreamsKotlin}. Also extracting bit values only by bit shifting operations instead of relying on an external library for handling the binary stream might improve reading speed. Another potential improvement in decoding speed could be achieved by reconstructing in memory and just write the finished file to disk. The main reason for the margin between encoding and decoding speed is most likely the multiple writing to the output file, since each bit position has to be decoded separately resulting in up to 8 write accesses to a single byte.
	 This could easily be resolved by first reconstructing in memory and only writing the file to disk once.
	 
	 \begin{table}[t!]
	 	\centering
	 	\begin{tabular}{l|r|r|r|r|r|r|r|r|r}	
	 		file & \multicolumn{1}{c|}{original}  & \multicolumn{4}{c|}{\texttt{ZIP}} & \multicolumn{4}{c}{proposed algorithm} \\ 
	 		 & \multicolumn{1}{c|}{size [kB]} & \multicolumn{1}{c|}{encode} & \multicolumn{1}{c|}{memory} & \multicolumn{1}{c|}{decode} & \multicolumn{1}{c|}{memory} & \multicolumn{1}{c|}{encode} & \multicolumn{1}{c|}{memory} & \multicolumn{1}{c|}{decode} & \multicolumn{1}{c}{memory} \\
	 		&  &  time [s] & peak [kB] & time [s] & peak [kB] & time [s] & peak [MB] & time [s] & peak [MB] \\
	 		
	 		\hline
	 		dickens & 9956 & 0.61 & 29.0 & 0.072 & 95.8 & 52.2 & 464 & 192 & 460 \\
	 		mozilla & 50020 & 1.99 & 29.0 &  0.331 & 98.6 & 191.1 & 1807 & 1080 & 1267 \\
	 		mr & 9740 & 0.52 & 29.0 & 0.070 & 99.3 & 49.8 & 456 & 163 & 459 \\
	 		nci & 32768 & 0.46 & 29.0 & 0.136 & 95.6 & 162 & 447 & 414 & 615 \\
	 		ooffice & 6008 & 0.31 & 29.0 & 0.054 & 98.0 & 31.9 & 446 & 145 & 464 \\
	 		osdb & 9852 & 0.32 & 29.0 & 0.069 & 97.6 & 49.9 & 367 & 223 & 551 \\
	 		reymont & 6472 & 0.37 & 29.1 & 0.045 & 97.9 & 32.7 & 476 & 119 & 392 \\
	 		samba & 21100 & 0.54 & 25.0 & 0.118 & 99.0 & 107 & 529 & 411 & 558 \\
	 		sao & 7084 & 0.41 & 25.0 & 0.058 & 93.5 & 37.8 & 461 & 184 & 459 \\
	 		webster & 40488 & 1.47 & 25.0 & 0.250 & 95.9 & 180.3 & 725 & 726 & 636 \\
	 		xml & 5220 & 0.09 & 25.0 & 0.023 & 97.3 & 28.2 & 284 & 154 & 354 \\
	 		xray & 8276 & 0.35 & 25.0 & 0.071 & 95.9 & 47.2 & 301 & 194 & 455 \\
	 		\hline
	 		all files & 206984 & 7.43 & 29.1 & 1.297 & 99.3 & 970 & 1807 & 4005 & 1267 \\
	 		\multicolumn{2}{l|}{$\varnothing$ values per file} & 0.61 & 27.3 & 0.108 & 97.0 & 80.8 & 563 & 333 &556
	 	\end{tabular}
	 	\caption{Runtime and memory peak of \texttt{ZIP v3.0} in comparison with the proposed algorithm, compressing and decompressing the Silesia Corpus}
	 	\label{tab:results_silesia_benchmark}
	 \end{table}

\section{Conclusions and future work}
	In conclusion, we demonstrated that with the help of different preprocessing steps and a different encoding technique, RLE can achieve compression results comparable to modern methods. 
	Not only is there a reasonable compression for every file in the different corpora containing a huge variety of data-types, files highly suited for the original proposed RLE were compressed even better.
	The relative file size after compression of our RLE based technique is with 42.34\% on average on files in the Canterbury Corpus only a few percent points behind daily used algorithms, e.g. \texttt{gzip} with 31.8\% or \texttt{ZIP} with 32.67\% and even slightly better than \texttt{compress} with 43.21\%.
	On raw image files like .PGM, .PPM, or .DNG, where a potential compression is desired to be lossless, our algorithm even achieves significantly better compression ratios than \texttt{ZIP}. 
	Despite the discussed potential for improvement, our implementations demonstrates the improvement of applicability of RLE to arbitrary input data by our discussed preprocessing steps.

	One interesting approach not performed in this scope is the encoding of Huffman codes after a byte-wise RLE instead of a vertical RLE. It was assumed to perform worse than the vertical encoding because there has to be one code for every combination of runs and values, thus very long average Huffman codes are expected. Another idea is the substitution of Huffman encoding by another, more sophisticated method like Asymmetric Numeral Systems~\cite{DBLP:journals/corr/Duda13}. This would most likely further improve compression results at the expense of slower computation. Furthermore we plan to implement a more efficient version of the algorithm with the help of the recently introduced high-level language abstractions framework for data compression by Ray et al.~\cite{DSL-2021}.

\textbf{Acknowledgment:}
The second author is supported by Deutsche Forschungsgemeinschaft project  
FE 560/9-1.

\bibliographystyle{plain}
\bibliography{refs}

\end{document}